\documentstyle[amssymb,prl,aps,epsfig]{revtex}

\begin{document}

\twocolumn[\hsize\textwidth\columnwidth\hsize\csname@twocolumnfalse\endcsname

\title{Realization of Bose-Einstein condensates in lower dimensions}
\author{A. G\"orlitz, J.~M. Vogels, A.~E. Leanhardt, C. Raman, T.~L. Gustavson, J.~R. Abo-Shaeer,
A.~P. Chikkatur, S. Gupta, S. Inouye, T.~P. Rosenband, D.~E.
Pritchard, W. Ketterle}
\address{Department of Physics and Research Laboratory of Electronics, \\
Massachusetts Institute of Technology, Cambridge, MA 02139}

\date{\today }
\maketitle
\begin{abstract}
Bose-Einstein condensates of sodium atoms have been prepared in
optical and magnetic traps in which the energy-level spacing in
one or two dimensions exceeds the interaction energy between
atoms, realizing condensates of lower dimensionality. The
cross-over into two-dimensional and one-dimensional condensates
was observed by a change in aspect ratio and saturation of the
release energy when the number of trapped atoms was reduced.
\end{abstract}
 \vskip1pc ]

New physics can be explored when the hierarchy of physical
parameters changes. This is evident in dilute gases, where the
onset of Bose-Einstein condensation occurs when the thermal
deBroglie wavelength becomes longer than the average distance
between atoms. Dilute-gas condensates of density $n $ in
axially-symmetric traps are characterized by four length scales:
Their radius $R_{\perp }$, their axial half-length $R_{z}$, the scattering length $%
a$ which parameterizes the strength of the two-body interaction,
and the healing length $\xi =(4\pi n a)^{-1/2}$. In almost all
experiments on Bose-Einstein condensates, both the radius and
length are determined by the interaction between the atoms and
thus, $R_{\perp },R_{z} \gg \xi \gg a$. In this regime, a BEC is
three-dimensional and is well-described by the so-called
Thomas-Fermi approximation  \cite{dalf99rmp}. A qualitatively
different behavior of a BEC is expected when the  healing length
is larger than either $R_{\perp}$ or $R_{z}$ since then the
condensate becomes restricted to one or two dimensions,
respectively. New phenomena that may be observed in this regime
are for example quasi-condensates
\cite{kaga87,petr00BEC2D,petr00BEC1D} and  a Tonk's gas of
impenetrable bosons \cite{petr00BEC1D,tonk36,olsh98}.

In this Letter, we report the experimental realization of
cigar-shaped one-dimensional condensates with $R_{z}>\xi >R_{\perp
}$ and disk-shaped two-dimensional condensates with $ R_{\perp
}>\xi
>R_{z}$. The cross-over from 3D to 1D or 2D
 was explored by reducing the number of atoms in
condensates which were trapped in highly elongated magnetic traps
(1D) and disk-shaped optical traps (2D) and measuring the release
energy. In harmonic traps, lower dimensionality  is reached when
$\mu_{3D} = 4\pi \hbar ^2 a\ n/m < \hbar \omega_t$. Here,
$\omega_t$ is the trapping frequency in the tightly confining
dimension(s) and $\mu_{3D}$ is the interaction energy of a weakly
interacting BEC, which in 3D corresponds to the chemical
potential. Other experiments in which the interaction energy was
comparable to the level spacing of the confining potential include
condensates in one-dimensional optical lattices
\cite{orze01squeezed} and the cross-over to an ideal-gas (zero-D)
condensate \cite{holl97int}, both at relatively low numbers of
condensate atoms.

Naturally, the number of interacting atoms in a lower-dimensional
condensate is  limited. The peak interaction energy of a 3D
condensate of $N$ atoms with mass $m$ is given by $\mu_{3D}
=\hbar ^{2}/2m \,(15Na/l_{z}^{2}l_{\perp }^{4})^{2/5}$, where $
l_{{\perp},z} = (\hbar/m \omega_{{\perp},z})^{1/2}$  are the
oscillator lengths of the harmonic potential. The cross-over to
1D and 2D, defined by $\mu_{3D} = \hbar \omega_t$ or equivalently
$\xi = l_t$ occurs if the number of condensate atoms becomes

\begin{eqnarray}
\label{equ:1D_number}
 N_{1D} = \sqrt{\frac{32  \hbar}{225 m a^2}}
\sqrt{\frac{\omega_{\perp}}{\omega_z^2}} =
\sqrt{\frac{\omega_{\perp}}{\omega_z^2}} \times 7200
\sqrt{\rm\frac{rad}{s}} \,
\\
\label{equ:2D_number} N_{2D} = \sqrt{\frac{32  \hbar}{225 m a^2}}
\sqrt{\frac{\omega_z^3}{\omega_{\perp}^4}} =
\sqrt{\frac{\omega_z^3}{\omega_{\perp}^4}} \times 7200
\sqrt{\rm\frac{rad}{s}} \, ,
\end{eqnarray}

\noindent where we have used the scattering length ($a=2.75$\,nm)
and mass of $^{23}$Na atoms to derive the numerical factor. Our
traps feature extreme aspect ratios resulting in  $N_{1D}
> 10^4$ and $N_{2D} > 10^5$ while for most standard BEC traps the numbers are
significantly smaller.

For a 1D condensate the condition $\xi = l_t$  yields a  linear
density $\tilde{n}_{1D}\approx 1/4 a$, implying that the linear
density of a 1D condensate is limited to less than one atom per
scattering length independent of the radial confinement.
Therefore, tight transverse confinement, as may be achievable in
small magnetic waveguides \cite{micro} or hollow laser beam guides
\cite{donut}, is by itself not helpful to increase the number of
atoms in a 1D condensate. Large 1D numbers may only be achieved at
the expense of longer condensates or if the scattering length is
smaller.

In anisotropic traps, a primary indicator of crossing the
transition temperature for Bose-Einstein condensation is a sudden
change of the aspect ratio of the ballistically expanding cloud,
and an abrupt change in its energy. The transition to lower
dimensions is a smooth cross-over, but has similar indicators. In
the 3D Thomas-Fermi limit the degree of anisotropy of a BEC is
independent of the number $N$ of atoms, whereas in 1D and 2D, the
aspect ratio depends on $N$. Similarly, the release energy in 3D
depends on $N$ \cite{dalf99rmp} while in lower dimensions, it
saturates at the zero-point energy of the tightly confining
dimension(s), when $N$ becomes smaller.

A trapped 3D condensate has a parabolic shape and its radius and
half-length are given by $R_{\perp} = l_{\perp}(15 N a l_{\perp}/
l_{z}^2)^{1/5}$ and $R_z = l_z (15 N a l_z^3/ l_{\perp}^4)^{1/5}$
\cite{dalf99rmp}, resulting in an aspect ratio of $R_{\perp
}/R_{z}= l_{\perp}^2/l_z^2 = \omega _{z}/\omega _{\perp }$. When
the 2D regime is reached by reducing the atom number, the
condensate assumes a Gaussian shape with an rms width $\approx
l_{z}$ along the axial direction, but retains the parabolic shape
radially. The radius of a trapped 2D condensate decreases with $N$
as $R_{\perp }=(128/\pi )^{1/8}(Nal_{\perp }^{4}/l_{z})^{1/4}$
\cite{petr00BEC2D}. Similarly, the half-length of trapped 1D
condensates is $R_{z} =(3Nal_{z}^{4}/l_{\perp }^{2})^{1/3}$
\cite{petr00BEC1D}.

\begin{figure}
\epsfig{file=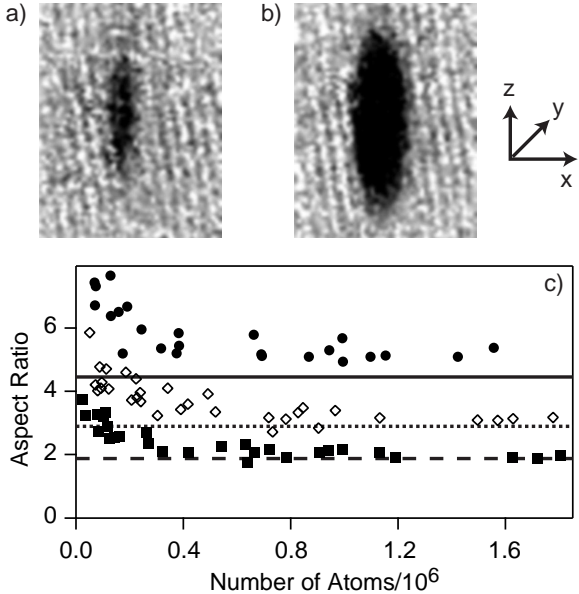,width=3in}
\newline
 \caption{Cross-over from 3D to 2D condensates observed in the change of the
 aspect ratio. Condensates were released from a disk-shaped optical trap and observed
 after 15\,ms
time-of-flight. a) (2D) condensate with $9\times10^4$ atoms b)
(3D) condensate with $8\times10^5$ atoms in a trap with vertical
trap frequency of $ \omega_z/2 \pi \approx 790$\,Hz. c) Aspect
ratio as a function of atom number for optical traps with vertical
trap frequencies of $1620$\,Hz (filled circles), $790$\,Hz (open
diamonds) and $450$\,Hz (filled squares). The lines indicate the
aspect ratios as expected for condensates in the 3D (Thomas-Fermi)
regime. We attribute discrepancies between expected and measured
aspect ratio for large numbers to the influence of
anharmonicities  on the measurement of the trap frequencies.}
\label{fig1}
\end{figure}

Our experiments in which one- and two-dimensional Bose-Einstein
condensates were realized were carried out in two different
experimental setups. For the study of condensates in a 2D
geometry, condensates of $\approx 10^{7}$ atoms were generated as
described in Refs. \cite{mewe96bec,kett99var} and transferred
into an optical trap \cite{stam98odt}. The optical trapping
potential was generated by focusing a 1064 nm laser into a light
sheet using cylindrical lenses, with the tight focus in the
vertical dimension to provide optimum support against gravity.
This resulted in typical trap frequencies of $\omega_z/2\pi =
790$\,Hz, $\omega_{\perp x}/2\pi = 30$\,Hz and $\omega_{\perp
y}/2\pi = 10$\,Hz for a laser power of $\approx 600\,${\rm mW}.
The transfer from the magnetic trap into the optical trap was
accomplished by turning on the trapping light field and turning
down the magnetic trapping potential resulting in a transfer
efficiency of more than 50\%. The depth of the optical potential
and the trap frequencies could be easily varied by changing the
power of the trapping beam.

To observe the transition from the 3D Thomas-Fermi regime into a
2D situation, we have adjusted the number of condensate atoms
between $2\times 10^{6}$ and $3\times 10^{4}$ by exposing the
optically trapped BEC to a thermal sodium atomic beam. Condensates
were detected by suddenly releasing the atoms from the trap and
taking absorption images after $15$\,{\rm ms} free expansion. The
condensates dropped by $ 15\,\mu$m during the $100\,\mu$s imaging
time, which is less than $10\%$ of the measured length of our
shortest condensates. During expansion, the interaction energy is
converted  almost exclusively into kinetic energy in the tightly
confining vertical direction. Thus, the vertical length of the
condensate expands quickly while the horizontal dimensions almost
retain their size in the trap. The aspect ratio changes
dramatically, and for the chosen time-of-flight the axial
dimension  is even larger than the radial one. The aspect ratio
was determined by fitting the condensate with a parabolic
Thomas-Fermi distribution which is exact in the large-number
limit. For small numbers the vertical condensate shape approaches
a Gaussian, but  we also used the parabolic fitting function to
avoid any bias. A parabolic fitting function may underestimate
the rms width of a Gaussian by at most $20 \%$.

\begin{figure}
\epsfig{file=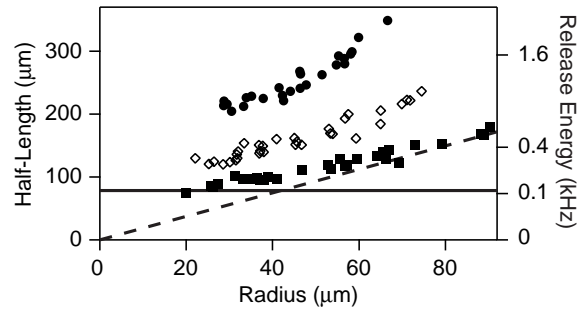, width=3in}
\newline
 \caption{Saturation of the half-length and correspondingly the release energy at
 the cross-over from 3D to 2D for the same optical traps as in
 Fig.\,\ref{fig1}. The dashed line represents the expected behavior of a purely 3D condensate
and the solid line the expected saturation level of the release
energy for the weakest trap (filled squares). } \label{fig2}
\end{figure}

Figures \,\ref{fig1}a,b show how the expanded condensate became
more elongated for small atom numbers, clear evidence for
reaching 2D. In Fig.\,\ref{fig1}c the aspect ratio of condensates
in three different optical traps is shown.
 For all our trap
parameters, the aspect ratio approaches a constant value for large
atom numbers, which can be calculated  using the results of
\cite{cast96} for the Thomas-Fermi limit. The increase of the
aspect ratio for small atom numbers is due to clamping of the
vertical length of the condensate because of saturation of the
release energy while the width shrinks further. For the weakest
trap, using Eq.\,\ref{equ:2D_number}, $N_{2D} = 2.9 \times 10^5$ ,
while we could observe condensates with as low as $ N_{2D}/10$.

The saturation of the mean release energy per particle at the
kinetic part of the zero-point energy of the trap becomes obvious
if the half-length of the expanded condensates is plotted versus
the radius (see Fig.\,\ref{fig2}). For a long enough
time-of-flight, the mean release energy is simply proportional to
the square of the measured half-length and is  given by $E_{rel}
= m R_z(t)^2/14 t^2$. In all our traps the half-length appears to
saturate at a value which corresponds to a release energy close
to $\hbar \omega_z/4$ which is the vertical kinetic zero-point
energy.

\begin{figure}
\epsfig{file=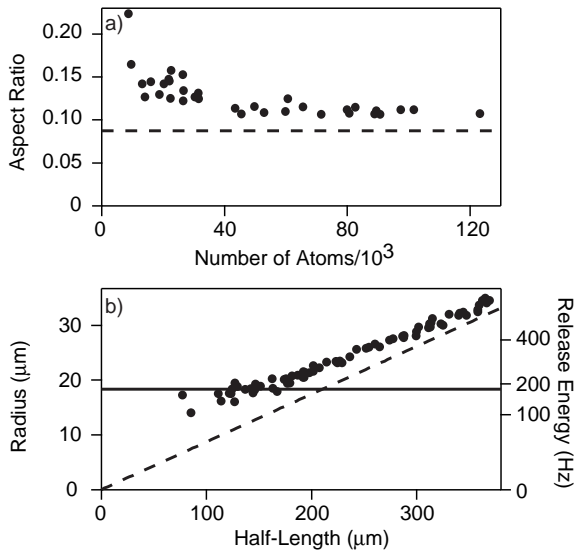,width=3in}
\newline
 \caption{Crossover from 3D to 1D condensates.  a)  Aspect ratio in
time-of-flight images versus the number of atoms.  b) Radial
versus axial size for the same time-of-flight data, showing
saturation of the release energy at close to the expected value
(solid line). The dashed line represents the behavior of a 3D
condensate. Note, that the use of a parabolic fitting function
slightly underestimates the release energy in the 1D regime. }
\label{fig3}
\end{figure}

The 1D experiments were carried out in a Ioffe-Pritchard type
magnetic trap with radial and axial trapping frequencies of
$\omega_{\perp}/2\pi = 360$\,Hz and $\omega_z/2\pi = 3.5$\,Hz
\cite{onof00}. We obtained an extreme aspect ratio of $\sim 100$
by reducing the axial confinement during the final evaporation
stage. As in the 2D case, the number of atoms in the condensate
was lowered by exposing the gas to a thermal sodium beam,
followed by a reequilibration time of $15$\,s. A radio-frequency
shield \cite{kett99var} limited the trap depth to $20-40$ {\rm
kHz} ($1-2\,\mu $K). We analyzed the cloud using absorption
imaging along one of the radial directions after a ballistic
expansion of $t = 4$ {\rm ms}.  In contrast to the 2D
experiments, the aspect ratio of the cloud was not yet inverted
after this short time-of-flight. The measured condensate sizes
were corrected for the finite imaging resolution of $5 \mu m$, a
correction of less than 10\%.

Similar to the 2D experiment, the cross-over to the 1D regime was
observed by a change of the aspect ratio when the number of atoms
was reduced (Fig.\ref{fig3}).  In the 3D limit, neglecting the
initial radial size and the axial expansion (both $<1\%$
corrections) the aspect ratio equals $\omega_z t$ independent of
$N$.  The deviation from this behavior below $\approx 5 \times
10^4$ atoms demonstrates the cross-over to 1D behavior.  At the
same time, the release energy saturated at $\hbar
\omega_{\perp}/2$, the zero-point kinetic energy of the trapping
potential (Fig.~\ref{fig3}b). For this trap, Eq.
\ref{equ:1D_number} yields $N_{1D}=1.6 \times 10^4$, while we
could observe condensates with as low as $N_{1D}/2$. In the 1D
geometry, the aspect ratio deviates from the 3D limit for  larger
ratios of $N/N_{1D}$ than in the 2D case, since the interaction
energy per particle at the cross-over to 1D is only approximately
half the kinetic zero-point energy while at the cross-over to 2D,
it is roughly equal to the kinetic zero-point energy.

 So far, we have discussed only the condensate and its cross-over
from the Thomas-Fermi regime into a lower-dimensional situation.
Here, we will briefly address the thermal component and finite
temperature effects. The finite trap depth $U_{{\rm trap}}$
 provides constant evaporative cooling which counteracts any residual
heating and stabilizes the temperature. The equilibrium
temperature in both the 1D and 2D experiment can be estimated to
be a constant fraction $1/\eta$ of the trap depth $ U_{\rm trap}$.
For a quantum saturated thermal cloud ($T<T_{c}$), the number of
thermal atoms $N_{{\rm th}}$ can be approximated by simply
counting the number of states with an energy below $k_B T$, which
results in $N_{{\rm th}}\approx (U_{{\rm trap}}/\eta\hbar
)^{3}/\omega _{\perp }^{2}\omega _{z}$. This estimate assumes that
the thermal cloud is still three-dimensional ($k_{B}T>\hslash
\omega _{{\perp},z}$), in agreement with the situation in the
experiment. To be able to discern the condensate from the thermal
cloud, $N_{{\rm th}}$ should not be much larger than $N_{1D}$
($N_{2D}$). This simple argument implies that for our 1D trap
where $N_{1D}\approx 1.6\times 10^4$, the maximum allowed trap
depth would be $60$\,kHz assuming a typical $\eta =10$ and a
minimum condensate fraction of 10\%. This is in fair agreement
with our experimental observations.

In a magnetic trap, the trap depth can be adjusted independently
of the trap frequencies using an rf shield. In an optical trap
created using a single Gaussian focus, the trap frequencies are
proportional to the square root of the trap depth. Thus, tighter
optical traps can store more thermal atoms, yet
Eq.\,\ref{equ:2D_number} implies that $N_{2D}$ is lower for
tighter traps. Therefore, tight single-focus optical traps are
less suitable for the observation of lower-dimensional
condensates. Experimentally, we have observed that for weaker
traps we could penetrate further into the 2D regime (see
Fig.\,\ref{fig1}c) than for tighter ones, which is consistent
with the considerations above.

In our experiments, the thermal cloud is always three-dimensional,
which necessarily requires that the critical temperature $T_c$ for
Bose condensation is also larger than the energy level spacing of
the trap, i.e. $k_B T_c > k_{B}T > \hslash \omega _{t}$. A new
physical regime in which the condensation process could be studied
in lower dimensions would be reached if thermal excitations freeze
out before a BEC forms. This requires that the total number of
atoms $N$ be smaller than the number of states with an energy
smaller than $\hbar \omega_t$. Thus, in 1D, the number of atoms
may not exceed the aspect ratio of the trap, i.e. $N < A=\omega
_{t}/\omega _{w}$, where $\omega_w$ is the trapping frequency in
the weakly confining direction, while in 2D the relevant criterion
is $N < A^2$. For the traps used in our experiments, this implies
$N<100$ (1D) or $N<2500$ (2D), which, at least for 2D, seems to be
within experimental reach. The physics of Bose-condensation in 2D
has been discussed in several publications. It is not yet clear
under what circumstances one will observe quasi-condensates
\cite{kaga87,petr00BEC2D} or the Kosterlitz-Thouless transition
\cite{mull00}. A full understanding of the observation of 2D
quantum degeneracy of spin-polarized hydrogen \cite{safo98prl} is
still lacking, and controlled experiments with dilute gases could
give useful insights. In a 1D geometry a related effect that could
be observed is a two-step condensation as discussed in
\cite{kett96}.

Lower-dimensional condensates, as prepared in our experiments,
offer many opportunities for further scientific studies.
Topological excitations such as solitons (in 1D) and vortices (in
2D) should be much more stable than in 3D, where solitons suffer
from kink instabilities and vortices can bend. The character and
spectrum of the collective excitations is expected to exhibit a
qualtitative change in lower dimensions \cite{ho99jltp}.

Another area of significant theoretical interest is the study of
phase fluctuations \cite{kaga87,petr00BEC2D,petr00BEC1D} and its
relation to the formation of quasi-condensates, which locally
behave like ordinary condensates but do not have a globally
uniform phase. In 3D, the energy associated with such phase
fluctuations is usually higher than $k_B T_c$ (except for very
extreme aspect ratios), whereas in 1D and 2D it can be lower. It
might be possible to observe the phase fluctuations by observing a
broadening of the Doppler spectrum of the condensate \cite{bragg}.

An even more ambitious goal  is the observation of a Tonks gas in
a one-dimensional geometry \cite{tonk36,petr00BEC1D,olsh98}. At
zero temperature, such a gas of ``impenetrable bosons'' is
realized when the axial distance $1/\tilde{n}$ between atoms
exceeds the 1D healing length, $\xi = l_{\perp }/(2 \tilde{n}
a)^{1/2}$ \cite{petr00BEC1D}, resulting in
$\tilde{n}_{\rm{Tonks}}= 2 a/l_{\perp }^{2}$ as a condition for
the linear density. Thus, $\tilde{n}_{\rm{Tonks}}$ is $8
a^2/l_{\perp }^{2}$ times smaller than $\tilde{n}_{1D}$. In 2D,
the collision physics is severely altered only for $a/R_{z}>1$
\cite{petr00BEC2D}. Such regimes require much tighter confinement
than in our experiment and may be realized using optical lattices
or magnetic microtraps, or alternatively a larger scattering
length, for example near a Feshbach resonance.

In conclusion, we have prepared lower-dimensional condensates in
optical and magnetic traps. The cross-over into lower-dimensional
behavior was indicated by a change in the aspect ratio of the
cloud and a saturation of the release energy as the condensate
number was lowered. Due to the extreme geometries of our traps the
number of atoms at the cross-over to lower-dimesionality is rather
large ($> 10^5$ in the 2D case) which provides a good starting
point for the exploration of  phenomena which only occur in one or
two dimensions.

This research is supported by NSF, ONR, ARO, NASA, and the David
and Lucile Packard Foundation. A.~E.~L. and A.~P.~C. acknowledge
additional support by a fellowship from NSF and JSEP,
respectively.

\end{document}